# Post-war Civil War Propaganda Techniques and Media Spins in Nigeria and Journalism Practice

Bolu John Folayan, PhD[1], Olumide Samuel Ogunjobi[2,] Prosper Zannu, PhD[3] Taiwo Ajibolu Balofin[4]

[1]Department of Mass Communication, Joseph Ayo Babalola University, Ikeji Arakeji, Osun State

[2]Department of Mass Communication, Federal University, Oye Ekiti, Ekiti State

[3]Department of Mass Communication, Lagos State Polytechnic, Ikorodu, Lagos

[4]Department of Mass Communication, Adeleke University, Osogbo, Nigeria

[1]bolujohnfolayan@gmail.com, [2]ogunjobiolumidesam@gmail.com, [3]prosperzannu@yahoo.co.uk, and [4]africaneye1.oal.com,

**ABSTRACT**

In public relations and political communication, a spin is a form of propaganda achieved through knowingly presenting a biased interpretation of an event or issues. It is also the act of presenting narratives to influence public opinion about events, people or and ideas. In war time, various forms of spins are employed by antagonists to wear out the opponents and push their brigades to victory. During the Nigerian civil war, quite a number of these spins were dominant – for example GOWON (Go On With One Nigeria); "On Aburi We Stand", "O Le Ku Ija Ore". Post-war years presented different spins and fifty years after the war, different spins continue to push emerging narratives (e.g. "marginalization", "restructuring"). This paper investigates and analyzes the different propaganda techniques and spins in the narratives of the Nigerian civil in the past five years through a content analysis of three national newspapers: **The Nigerian Tribune**, **Daily Trust** and **Sun Newspapers**. Findings confirm that propaganda and spins are not limited to war time, but are actively deployed in peace time. This development places additional challenge on journalists to uphold the canons of balance, truth and fairness in reporting sensitive national issues. The authors extend postulations that propaganda techniques, generally considered to be limited to war situations, are increasingly being used in post-war situations. Specifically, they highlight that journalists are becoming more susceptible to propaganda spins and this could affect the level of their compliance to the ethics of journalism.

**Keywords**: Biafran War, Framing, Propaganda, Media Spins.

**INTRODUCTION**

The place of Communication and the media in Warfare

Communication – the sharing of information, ideas, knowledge and meanings – is basic to existence. It takes place in various forms and for various purposes. This can be between humans, between non-humans, between humans and machines, between machines, humans to humans but machine-mediated, etc. Some religious philosophers even argue that the dead humans do communicate. Thus, communication is a phenomenon that is applied in all fields of endeavor.

One area in which communication has demonstrated pervasive impact is conflict and warfare. (Folayan, 2011). One of the reasons for the occurrence of the Nigeria-Biafra war (also known as the Nigerian civil war) - and the temporary successes and failures on either sides during the war - was the effectiveness (or non-effectiveness) of communication. Though the remote causes of the war were religious, cultural, economic, political and ethnic, it has been noted that the foregoing were normal occurrences in political entities with diverse ethnic compositions and that the eventual breakdown of communication was one of the immediate causes of the war. (Akinyemi, 1972).

Soon after independence in 1960, political altercations between the mainly Muslim-North and the Christian-dominated East, and the political crises in the West, the country became sharply divided along regional lines. The military took over the reins of power in 1966 and months later a counter-coup took place. Both coups were had ethnic sentiments, especially with the killings of political leaders of Northern and Western origins and none of Eastern origin. (Ugochukwu, 1972). Major-General Johnson Aguiyi-Ironsi, became the head of state after the 1966 coup but was also killed in the counter-coup which heightened ethnic tensions. On May 30, 1967, the Igbo-majority province of the East declared the Republic of Biafra. The war began effectively after the final collapse of the "Aburi talks" in Ghana, a neigbouring African country.

*The Communication Media and Propaganda during the Nigeria-Biafra War*

The Biafran leaders constituted the Ministry of Information and set up a strong 'Propaganda Directorate'. (Ugochukwu, 1972). The major role of the Ministry and the Directorate was to manage the inflow and outflow of information within the military and between the military and Biafrans on one hand, and between Biafra and the Nigerian governments and other foreign governments and international stakeholders.

The key targets for Biafra's and Nigeria's propaganda were the international broadcast media (Voice of America , BBC World Service, , France-Inter, Radio-France International (RFI), Radio-Lausanne, Europe N.1,5, Radio-Brussels and Radio-Canada)., (Lewis, 1970).Ugochukwu has reported details of how both sides effectively used the broadcast media:

They gave details of military operations and tried to analyse rumours and clues about possible arms sales to the warring parties; they also reviewed the humanitarian situation, followed diplomatic progress and political declarations from various countries, and discussed the peace prospects… reports on the military and humanitarian fronts in order to evaluate potential changes in the reporting of the news and the difficulties faced by the journalists as the war progressed. News items in Spanish were aimed at Sao Tome while Hausa news targeted Federal troops that recruited mostly in northern Nigeria; Tiv, Idoma, Igala and Yoruba news targeted neighbouring southern Nigerian states. Biafran news bulletins were mostly dispatched through the Geneva office of Markpress, a public relations firm owned by American adman H. William Bernhardt. According to the then *Times* correspondent, between January and August 1968, there were more than 250 of these press releases [http://www.time.com/time/magazine/article/0,9171,838607-1,00.html]. According to Jung (2007), in the first half of 1968, Markpress had arranged air passages into Biafra for more than 70 newsmen from every western European nation, and transmitted eyewitness reports to their publications.

(Ugochukwu, 1972)

The print media were more effectively used by the Nigerian government but according to Stafford (1984) the print media had also adopted the same pattern of its reporting of news from the war front: Press releases from the two sides were so distorted that the *New York Times*, for example, ran adjacent Biafran and Nigerian sources stories.

*Rationale for the Study*

This study arose from the need to examine some of the major propaganda techniques used in managing post-war conflicts. Much has been written about the role of the media in the Biafran War since 1970; including, studies on the media involvement in the conflict. Classic works on this include those by Akinyemi (1979), Standford (1984) and Uche, (1989).

Some of these studies led to accusations of bias on the part of the media, largely because inadequate efforts were focused, in the studies, on the propaganda elements in the reports published by both the electronic and print media. Much of what was reported were works of public relations and propaganda experts and because of the war situation, the media houses did not have the capacity to confirm the authenticity of what they published. Even in post-war times, propaganda communication is difficult to detect; how much more in war situations. For example, Uche (1989) noted that the press reported counter-claims of military victories long before they were fact. According to him, one of the most famous examples of this disinformation was the front page of the London *Times* that claimed that Federal troops had been cleared of the killing of four European relief workers in Okigwe on 30 September 1968. (Uche, 1989).

Many players in the war, such as military tacticians also published aspects of these propaganda in their various accounts, but only systematic or scholarly investigations could properly unveil such carefully-planned and executed manipulated communications; especially their intents, purposes and techniques.

This investigation, therefore, takes a different approach by looking beyond the surface of what the media reported; making deeper inquiry into the propaganda elements that might have nurtured what the media reported, after the war. As has been established, propaganda is not limited to conflict or wartime: It takes place before, during and after war and conflicts. Thus, this investigation will help in providing insight to understanding war and conflict-related information published in the media. It will also help journalists to make better interpretations and analyses of conflict-related stories, press releases, features and other forms of strategic communication related to the Nigeria-Biafran war.

*Research Questions*

The investigation sought answers to three questions:

1. What propaganda techniques on Biafran war-related issues were deployed in the selected media during the period under review?
2. What media spins were deployed on issues related to the Nigerian civil war in the newspapers under review?
3. What were the major sources of post-war propaganda information in the newspapers under review?

**LITERATURE REVIEW**

*Conceptual Review*

Propaganda is the deliberate and systematic attempt to shape perceptions, manipulate cognitions and direct behaviour to achieve a response that furthers the desired intent of the propagandist. (Jowett & O'Donnel, 2018). It is persuasion that is deceptive, sometimes frightening, unscrupulous and false but sometimes, as Oladimeji (2015) has noted, it can be "clean word rendered dirty through its application to denounce enemy tricks in war fare" (Columbiaonline.com, 2015). In broader language, it may be defined as "systematic manipulation of public opinion, generally by the use of symbols, such as flags, monuments, oratory, and publications."(Oladimeji, 2015, p.45) Modern propaganda in non-war time is distinct from other forms of communication in that it is consciously and deliberately used to influence group attitudes. Thus, almost any attempt to sway public opinion, including lobbying, commercial advertising and missionary work, can be broadly construed as propaganda.

The origin of propaganda is often traced to 1622 when the Roman Catholic Church set up the *Sacra Congregatio de Propaganda fide,* the branch of the Church responsible for propagating the faithy. In this original meaning, propaganda, though truthful, was also covert: It persuaded people without seeming to do so. In the course of its application to various human endeavors, especially to wars and abnormal situations, the term earned its negative connotation,

<u>Black, grey and white propaganda</u>

Distinction is often made of three kinds of propaganda in modern times, based on the intent, intensity and method or style of the propagandist. *Black Propaganda* is deliberate and strategic transmission of lies, according to Baran and Davis .Although it is also used in peacetime, this kind of propaganda is most commonly used in crises situations such as wars.  *White Propaganda* has to do with the propagation of truth. It is commonly used in advertising and marketing communications, where a company tells the consumers all the beautiful things about a product but is silent on the negative aspects of the product. According to Lamprecht, white propaganda is simply putting out full facts of a story and sometimes by preventing people from being confronted with opposing points view. (Lamprecht, 2001).  *Grey Propaganda* is a mixture of truth with carefully-selected lies. At its finest, grey propaganda contains verifiable facts with credible sounding lies – which cannot be verified - weaved into it. Those who read or hear a grey propaganda tend to believe it because they can verify part of the story and that makes them believe the entire story (including the part they cannot verify). (Buhler, 1968). Experts consider grey propaganda as the most potent of all three types of propaganda. (Lamprecht, 2001).  This may be due to the brazen lies contained in black propaganda and the heavy impacts obtained when truth is mixed with falsehood. (Klaehn , Broudy, Fucha, & Godler, 2018).

Lampretch (2001) gives other reasons for the effectiveness of grey propaganda include the following:

- Message does not have to be untrue

- Technique often credible because it involves two-way communication (unlike black propaganda which is usually one-way)
- The source is usually hidden, and when revealed, it is often 'transferred'
- Presents misleading information in a more insidious manner than white propaganda
- Deliberately evokes a strong emotion, especially by suggesting illogical (non-intuitive) relationships between concepts
- Avoids the distinctively biased or one-sided rhetoric and works by presenting a contrived premise for an argument as if it were a universally accepted so that the audience naturally assumes it to be true.

Propaganda and war

Communication plays a crucial role in successful military leadership. A military commander to be an effective communicator and must be versatile in interpersonal relationships, consciousness of nonverbal messages and emotional relationships. (Lewiska, 2016).

Veterans in military warfare agree that warfare communication often holds the keys to victory.

Communicating with your allies while knowing where your enemies are is one of the most crucial parts in war. After all, in military ranks the common saying is that, "knowing is half the battle." Military communication has evolved throughout the ages from flaming arrows, drum beats, smoke signals, messenger pigeons, to modern satellite enabled communication devices (nmmc.com, 2018)

During the First World War, (WWI) communication technology was changing very quickly. For the first time, much of the world was using electricity, and this new source of power was utilized for communication in the form of telegraphs, telephones, signal lamps, As military commander uses to enhance (or weaken) the strength of relations between him/her and soldiers. If he/she is a proficient "communicator", conscious of power of the body language and manner of transmitting the contents, his/her chances of becoming a real leader significantly increase. Most of military communication is inherently propaganda-oriented and this could be *concealed propaganda* or *revealed propaganda*. (Tewsbury & Scheufele, 2010)

Spins

In the context of communication, spinning is the assemblage of facts and shaping of information to support a particular story; hence it can be positive and negative. In propaganda, to spin an issue is to communicate it in a way that changes the way people are likely to perceive it. A political spokesperson hired to promote favourable or pre-conceived interpretation of events to journalists (hence to the audience) is termed "the spin doctor" and the impact achieved is known as "media spins" (Scheufele, 2010). The description by former American president, Bill Clinton, of his wrong relationship with an intern as "inappropriate relationship" is a classic example of "media spin".

*Theoretical Review*

The central theoretical framework for this study is the Framing Theory. Framing comprises of a set of conceptual or theoretical perspectives on how people or groups (are made to) perceive, locate, identify and construct reality. In this regard, frames could be in form of how people think of or interpret issues (mental frames) or in form of manipulations that take place between individuals and groups during the communication process (communication frames). In journalism and advertising, framing can be used to shape how the listener, viewer or reader perceives an issue and public opinion is formed. The resulting perception is therefore known as "framing effect". (Scheufele, 2010; Lukmantoro, T. 2019).

Framing theory, often credited to be propounded by Goffman in his book, "Frame Analysis" ,has two broad perspectives - the *sociological* and *psychological* approaches.(Goffman, 1974)

Psychological framing assumes that people make individual judgment and perception within certain frames of reference; therefore it is possible to set up" situations in which appraisal of a social situation will be judged by such people.  On the other hand, sociological framing suggests that society operates on social categories and when issues are placed differently within these categories, interpretation and evaluation would be done in context of the categories. (Fairhust, 1996l; Drukkman, 2001).

How an issue is packaged in the news media can remarkably affect how people understand or react to such an issue. Framing may be done by simply providing information. But can also be done in more sophisticated forms, as Tewksbury and Scheufele (2010:p.19) have explained - through arguments, information, symbols, metaphors and images… that affect how people understand issues and at their core, issue packages have a frame – a central organizing idea or story line that provides meaning to an unfolding strip of events. (Tewksbury & Scheufelle, 2010). It can also be done to achieve persuasive effects. From the persuasive angle, the framing process is laced with elements that tend to influence attitudes in a predictable direction. .

Framing has meta-theoretical similarity with the Agenda Setting Theory (the latter describes how the media helps to know what to think *about,* not necessarily what they *think*), but it is different in that *framing* explains the processes through which the agenda is set. Frames are abstractions that work to structure communication and understanding and has been referred to as "the second-level agenda setting". When editors and sub-editors write or re-write stories, place them in positions of prominence or obscurity in media space, they are actually framing those stories and such framing do often influence the perception of the story by the audience. (Borah, 2011).

Fairhurst and Sarr (1996) have identified seven major means of framing in mass and interpersonal communication as follows:

- Metaphor: to frame a conceptual idea through comparison to something else.
- Tradition (rituals, ceremonies): Cultural mores than imbue significance in the mundane, closely tied to artifacts.
- Stories (myths, legends): To frame a topic via narrative in a vivid and memorable way.
- Contrast: To describe an object in terms of what it is not
- Spin: To present a concept in such a way as to convey a value judgment (positive or negative) in a hidden form.
- Slogan, jargon, catch phrase: To frame an object with a catchy phrase to make it more memorable.
- Artifact: Objects with intrinsic symbolic value – a visual /cultural phenomenon that holds more meaning than the object itself.

Related Studies

Although numerous studies have been done on the art of propaganda regarding the Nigerian civil war, not many of them actually dealt with the frames and schemes involved, especially from the perspective of the print media, after the war. Studies by Akinyemi (1979), Bach (1982), Stafford (1984), and Uche (1989) are good examples in this regard, although they mostly concentrated on the broadcast media. A very recent study by Abati, Onifade and Fasanu (2019) explored the perennial tension among the major ethnic groups and reforms in Nigeria and the interplay between ethnicity and politics in the media in relation to the Igbo, among others. Findings show that the ethnicity of the publishers as well as the geographical location of the publication affected the framing of reports. The researchers found that journalists working for the newspapers studied were biased toward sectional interests in their narratives.

**MATERIALS AND METHODS**

The method of gathering data for this investigation was the Content Analysis. Three national newspapers were selected based on their perceived regional appeals (even though they are national newspapers. These are:

1. *Nigerian Tribune* (more popular in the West)
2. *The Sun* (more popular in the East)
3. *Daily Trust* (more popular in the North)

Scope of study was five years (2015-2019), giving a population of 1,825 copies for each of the three selected newspapers per year (i.e. total of 5,475 newspapers for five years). Two copies of each news media were selected per week through systematic random sampling to generate a sample size of 104 copies per year, (or 520 copies of each newspaper over five years). This means that a total of 1,560 newspapers was eventually selected and content-analyzed. A coding sheet was designed based on the Research Questions to generate data from news stories, features, cartoons, photographs, photo-captions, editorials and letters to the editor.

**FINDINGS AND DISCUSSION**

*Propaganda techniques on Biafran war-related issues deployed through the Newspapers*

TABLE 1 – PROPAGANDA TECHNIQUES DEPLOYED IN THE PERIOD UNDER REVIEW

| SN | PROPAGANDA TECHNIQUE | PROPAGANDA TYPE | FREQUENCY | PERCENTAGE |
|---|---|---|---|---|
| 1 | Cooked Intelligence | Black | 119 | 5.0% |
| 2 | Name Calling | Grey | 196 | 8.3% |
| 3 | Word games/glittering generality | Grey | 337 | 14.2% |
| 4 | False connections | Black | 71 | 3.0% |
| 5 | Fear Appeal | Grey | 361 | 15.2% |
| 6 | Logical fallacy/faulty logic | Black | 308 | 13.0% |
| 7 | Unwarranted extrapolation | Grey | 47 | 2.0% |
| 8 | Stereotype | Grey | 24 | 1.0% |
| 9 | Transfer | Grey | 157 | 6.6% |
| 10 | Misleading headlines | Grey | 47 | 2.0% |
| 11 | Disinformation | White | 237 | 10.0% |
| 12 | Direct order | White | 5 | 0.2% |
| 13 | Association | White | 379 | 16.0% |
| 14 | Use of slogans | White | 24 | 1.0% |
| 15. | Plain folks | White | 59 | 2.5% |
|  | TOTAL |  | 2,371 | 100% |

At least fifteen (15) propaganda techniques were used in the various publications related to the civil war and self-determination of the Igbo people in the period under review. The most common techniques were the *grey* and *white* propaganda techniques (roughly accounting for 40 per cent each of the techniques deployed. *Black propaganda* was the least deployed of the three main propaganda types. (See Table 1). This agrees with Jewett and O'Donnor (2018) that black propaganda is not commonly used in peace time for agitations.

Within the *grey techniques,* fear appeals, glittering generalities and word games were most common; whereas for *white propaganda*, disinformation and association were most common. *Black propaganda techniques*, featured crooked intelligence and logical fallacy more than other techniques. The Police (pro-Federal Government) and the IPOB leadership often engaged in the use of "direct order" and "unwarranted extrapolation", while the IPOB leader, Nnamdi Kanu (who engaged in hide-and-seek games culminating in the jumping of bail granted by the Federal authorities) propagated lots of "disinformation" (non-deliberate misleading information) and "misinformation" (deliberate publishing of false information). He often made wild claims about his health and number of "killings" of his followers, by the Nigerian Police, using "association" (for example, claiming that "all" Igbo in the Diaspora had met with him to take certain decisions. The high number of "misleading headlines" could be a direct result of false and conflicting stories planted in the media. Most of the reports were not verifiable or could not be verified at the time of going to press.

*Media spins deployed on issues related to the Nigerian civil war in the Newspapers*

TABLE 2: MEDIA SPINS DEPLOYED IN THE PERIOD UNDER REVIEW

| SN | MEDIA SPINS | FREQUENCY | PERCENTAGE |
|---|---|---|---|
| 1 | Biafra/Biafrans | 76 | 3.2% |
| 2 | Marginalization | 71 | 3.0% |
| *3* | *Restructuring* | *43* | *1.8%* |
| *4* | *Tribalism* | *9* | *0.4%* |
| *5* | *Ethnic militia* | *19* | *0.8* |
| *6* | *Self-determination* | *13* | *0.5* |
| *7* | *Indigenous People of Biafra (IPOB)* | *341* | *14.4* |
| *8* | *Ndigbo* | *59* | *2.5* |
| *9* | *Revolution* | *59* | *2.5* |
| *10* | *Break-up* | *41* | *1.7* |
| *11* | *Nigerian civil war* | *106* | *4.5* |
| *12* | *Igbo nation* | *23* | *1.0* |
| *13* | *IPOD community radio* | *24* | *1.0* |
| *14* | *Agitation* | *61* | *2.6* |
| *15* | *Dialogue* | *63* | *2.7* |
| *16* | *Killings* | *106* | *4.5* |
| *17* | *Attacks* | *93* | *3.9* |
| *18* | *Igbo leaders* | *81* | *3.4* |
| *19* | *Igboland* | *43* | *1.8* |
| *20* | *Barbaric shootings* | *73* | *3.1* |
| *21* | *Incessant clashes* | *73* | *3.1* |
| *22* | *Igbo socio-cultural organization/Ohaneze* | *101* | *4.3* |
| *23* | *Yoruba socio-cultural organization/Afenifere* | *54* | *2.3* |
| *24* | *Arewa Consultative Forum/Youths* | *79* | *3.3* |
| *25* | *Unarmed people* | *4* | *0.2* |
| *26* | *Proscribed/outlawed (adjective)* | *45* | *1.9* |
| *27* | *Secession* | *37* | *1.6* |
| *28* | *Embrace peace* | *18* | *0.8* |
| *29* | *Foremost (adjective)* | *11* | *0.5* |
| *30* | *Unlawful assembly/association* | *63* | *2.7* |
| *31* | *Separatist movement* | *10* | *0.4* |
| *32* | *False allegations* | *36* | *1.5* |
| *33* | *Worldwide* | *15* | *0.6* |
| *34* | *Terrorist group* | *17* | *0.7* |
| *35* | *Declaration* | *29* | *1.2* |
| *36* | *Clash* | *60* | *2.5* |

| 37 | *Movement for the Actualization of the Sovereign State of Biafra, MASSOB* | *153* | *6.5* |
|---|---|---|---|
| 38 | *Compensate/compensation* | *6* | *0.3* |
| 39 | *Northerners/North* | *41* | *1.7* |
| 40 | *Easterners/East* | *19* | *0.8* |
| 41 | *Westerners/West* | *30* | *1.3* |
| 42 | *Condemned* | *66* | *2.8* |
|  | **TOTAL** | **2,371** | **100.0%** |

Forty-two (42) spin words were identified from the analysis of reports, features, news stories, photographs and editorials of the three newspapers sampled. Findings show that IPOB and MASSOB most commonly used, most likely for being the protagonists of the agitations. Although the expressions and words (listed in Table 2) have connotative or original meanings, reading through the context of the reports reveal denotative slants to suit the interests of the various stakeholders (Federal government, Police, IPOB and MASSOB and their different supporters). For example, the Federal Government often described IPOB as "illegal", "secessionist", "unlawful", "separatist", "terrorist" and "discredited" organization while the IPOB and MASSOB leaders described the police of "incessant killing" of their members. The leader of IPOB, Nnamdi Kanu engaged often in "association" – for instance, claiming that he had meeting "with all Igbo in the Diaspora", and describing IPOB as a "foremost" Igbo group. All stakeholders engaged in "false allegations" and counter-allegations in one form of the other as well as "misleading headlines". The case of "misleading headlines" may have resulted from the conflicting information feeds by various stakeholders and journalists are often not able to verify the authenticity of such claims due to production schedule, and especially when "grey propaganda" (mixture of truth and untruth) are dominant. The IPOB leader often engaged in unwarranted extrapolation and association to give the impression that the organization has massive grassroots support. On its part, the Federal Government used media spins to portray the IPOB and its leader as "illegal" but on the interests of Igbo ethnic group in general, the government engaged more in "White Propaganda" techniques, using softer expressions such as "Igbo are not marginalized", ""The East needs to integrate and embrace peace", etc.

*TABLE 3: MEDIA SPINS ACROSS NEWSPAPERS*

| *NEWSPAPER* | *NO OF SPINS* | *PERCENTAGE* |
|---|---|---|
| *Nigerian Tribune* | *745* | *31.4%* |
| *The Sun* | *1,034* | *43.6%* |
| *Daily Trust* | *592* | *25.0%* |
| *TOTAL* | *2,371* | *100.0%* |

The newspapers reinforced their "regional biases" in the volume of spins as earlier reported. (Abati, et al. 2019). *The Sun* newspaper, generally-perceived as pro-Igbo recorded higher volume of stories and spins. (Table 3). But as Table 4 shows, the publications were generally balanced. All three newspapers reported reasonably high number of positive (pro-Federation) as well as negative (anti-Federation) stories. *The Sun*, surprisingly had more Pro-Federation stories than both *Tribune* and *Daily Trust*. This may have occurred due to the higher volume of reportage (43.6 per cent) by *The Sun*.

*TABLE 4: DIRECTION OF MEDIA SPINS ACROSS NEWSPAPERS*

| Newspaper | Positive | Negative | Total |
|---|---|---|---|
| Tribune | 21% (499) | 10.3% (316) | 31% (815) |
| The Sun | 23% (554) | 20.7% (379) | 44% (933) |
| Daily Trust | 10% (230) | 15.0% (389) | 25% (619) |
| Total | 54% (1283) | 46% (1084) | 100.0% (2,367) |

*Major sources of post-war propaganda information in the Newspapers*

*TABLE 5: MAJOR SOURCES OF POST-WAR PROPAGANDA INFORMATION*

| Newspaper | Frequency | Percentage |
|---|---|---|
| Press statement | 421 | 17.8 |
| Opinion articles | 534 | 22.6 |
| News coverage | 435 | 18.4 |
| Interviews | 385 | 16.3 |
| Features | 592 | 25.0 |
| Total | 2,367 | 100% |

In terms of sources of stories, press statements and news reports accounted for the least volume (17.8 per cent and 18.4 per cent respectively) whereas most contents analyzed were opinions of members of the public (22.6 per cent). (See breakdown in Table 5). It should be noted, however, that in media-spins, "ghost writers" could be hired to make contributions, making it difficult to know the true sources of the information. Most propagandists often hide or disguise sources. Generally, however, news coverage and interviews account for over one-third of the post-war propaganda information gathered from the three newspapers during the period.

**CONCLUSION & RECOMMENDATIONS**

This study examined the contents three leading Nigerian newspapers with distinct regional popularity in terms of how they reported post-civil war news and discourse with a view to finding out how the stakeholders engendered support for their agitations. The findings confirm that manipulative communication is not limited to crises period or wartime. The study suggests that a major difference in media and propaganda in war time and post-war time is in technique. In wartime, *black propaganda* is more frequently deployed while in post-war time *grey propaganda* is more massively deployed. In the thick of this selfish communications, media personnel may be handicapped in being able to uphold the cannons of truth, balance, and fairness. Media editors' responsibility to report "all sides" of stories and features concerning the Nigerian civil war have therefore been compromised to an extent. This might not have been deliberate on the part of the editors, as most of the stories had skillfully embedded as slants in press releases and interviews.

The finding that even national newspapers were making regional focus with respect to the post-Biafra narrative further suggest that the spins and propaganda techniques have potentials to compromise objectivity in journalism. The editors need to be further trained to read between the lines regarding very sensitive issues such as the Nigerian civil war.